\documentclass[slac_one]{revtex4}
\usepackage{graphicx}
\usepackage{fancyhdr}
\usepackage{color}
\usepackage{amssymb,amsmath,array}
\begin{document}

\title{Matching of Resummed NLLA with Fixed NNLO for Event Shapes} 

%

\author{T.~Gehrmann, G.~Luisoni}
\affiliation{Institut for Theoretical Physics, Universit\"at
Z\"urich, CH-8057 Z\"urich, SWITZERLAND}
\author{H.~Stenzel}
\affiliation{Physikalisches Institut, Justus-Liebig Universit\"at
Giessen, D-35392 Giessen, GERMANY}

\begin{abstract}
We report work on the matching of the next-to-leading logarithmic
approximation (NLLA) onto the fixed next-to-next-to-leading order
(NNLO) calculation for event shape variables in electron-positron
annihilation. The correction of the combined NLLA+NNLO computation
in the three-jet region, relevant for precision phenomenology, is
small compared with pure NNLO or NLLA+NLO.
\end{abstract}

\maketitle

\thispagestyle{fancy}


\section{INTRODUCTION} 

Event shape distributions in $\mathrm{e^+e^-}$ annihilation
processes are very popular hadronic observables, mainly due to the
fact that they are well suited both for experimental measurement and
for theoretical description because many of them are infrared and
collinear safe. The main idea behind event shapes variables is to
parameterize the energy-momentum flow of an event, such that one can
smoothly describe its shape passing from pencil-like two-jet
configurations, which are a limiting case in event shapes, up to
multijet final states. Since the deviation from two-jet
configurations is proportional to the strong coupling constant
$\alpha_{s}$, the comparison of experimental measurements and
theoretical prediction permits to determine $\alpha_{s}$.

At LEP a set of six different event shape observables were measured
in great detail: thrust $T$ (which is substituted here by $\tau =
1-T$), heavy jet mass $\rho$, wide and total jet broadening $B_W$
and $B_T$, $C$-parameter and two-to-three-jet transition parameter
in the Durham algorithm $\mathrm{y}_3$.

Until very recently, the theoretical state-of-the-art description of
event shape distributions was based on the matching of the
NLLA~\cite{resumall} onto the NLO~\cite{ERT,event} calculation.
Using the newly available results of the NNLO\footnote{Recently an
inconsistency in the treatment of large-angle soft radiation was
discovered~\cite{weinzierlnew}. It is about to be corrected and it
should result in numerically minor changes to the NNLO coefficients
in the kinematical region of phenomenological studies here. The
corrections turn out to be significant only in the deep two-jet region,
e.g. $(1-T)\ll 0.05$.} corrections for the standard set of event
shapes~\cite{ourevent} introduced above, we computed the matching of
the resummed NLLA onto the fixed order NNLO.

\section{FIXED ORDER AND RESUMMED CALCULATIONS}

At NNLO the integrated cross section
\begin{equation}\label{Rfixed}
R\left(y,Q,\mu\right)\,\equiv\,\frac{1}{\sigma_{{\rm
had}}}\int_{0}^{y}\frac{d\sigma\left(x,Q,\mu\right)}{dx}dx,\nonumber
\end{equation}
has the following fixed-order expansion:
\begin{equation}\label{Rfixedexp}
R\left(y,Q,\mu\right)\,=\,1+\,\bar{\alpha}_{s}\left(\mu\right)\mathcal{A}\left(y\right)\,
+\,\bar{\alpha}_{s}^{2}\left(\mu\right)\mathcal{B}\left(y,x_\mu\right)\,+\,\bar{\alpha}_{s}^{3}\left(\mu\right)\mathcal{C}\left(y,x_\mu\right).\nonumber
\end{equation}
where $\bar\alpha_s = \alpha_s/(2\pi)$ and $x_\mu = \mu/Q$.
Approaching the two-jet region the infrared logarithms in the
coefficient functions becomes large, spoiling the convergence of the
perturbation expansion. The main contribution in this case comes
from the highest power of the logarithms which have to be resummed
to all orders. For suitable observables resummation leads to
exponentiation. At NLLA the resummed expression is
\begin{equation*}\label{eq:Rresummed}
R\left(y,Q,\mu\right)\,=\ \left(1+C_{1}\bar{\alpha}_{s}
\right)\,e^{\left(L\,g_{1}\left(\alpha_{s}L\right)
+g_{2}\left(\alpha_{s}L\right)\right)}\;,
\end{equation*}
where the function $g_{1}\left(\alpha_{s}L\right)$ contains all
leading-logarithms (LL), $g_{2}\left(\alpha_{s}L\right)$ all
next-to-leading-logarithms (NLL) and $\mu=Q$ is used. Terms beyond
NLL have been consistently omitted.
\begin{table}[t]
\centering\Large
\begin{tabular}{|l|c|c|c|c|c|c|}
       \hline
        ${\bar{\alpha}_{s}\mathcal{A}\left(y\right)}$ & $\color{blue}\bar{\alpha}_{s}L$ & $\color{red}\bar{\alpha}_{s}L^{2}$ &  &  &  &  \\\hline
        ${\bar{\alpha}_{s}^{2}\mathcal{B}\left(y,x_\mu\right)}$ & $\bar{\alpha}_{s}^{2}L$ & $\color{blue}\bar{\alpha}_{s}^{2}L^{2}$ & $\color{red}\bar{\alpha}_{s}^{2}L^{3}$ & $\color{green}\bar{\alpha}_{s}^{2}L^{4}$ &  & \\\hline
        ${\bar{\alpha}_{s}^{3}\mathcal{C}\left(y,x_\mu\right)}$ & $\bar{\alpha}_{s}^{3}L$ & $\bar{\alpha}_{s}^{3}L^{2}$ & $\color{blue}\bar{\alpha}_{s}^{3}L^{3}$ & $\color{red}\bar{\alpha}_{s}^{3}L^{4}$ & $\color{green}\bar{\alpha}_{s}^{3}L^{5}$ & $\color{green}\bar{\alpha}_{s}^{3}L^{6}$\\\hline
\end{tabular}
\caption{Powers of the logarithms present at different orders in
perturbation theory. The color highlights the different orders in
resummation: LL (red) and NLL (blue). The terms in green are
contained in the LL and NLL contributions and exponentiate trivially
with them.}\label{tab:logs}
\end{table}
The resummation functions $g_1(\alpha_s L)$ and $g_2(\alpha_s L)$
can be expanded as power series in $\bar{\alpha}_{s}L$
\begin{eqnarray}\label{eq:gexpand}
L\,g_{1}\left(\alpha_{s}L\right)&=&\,G_{12}L^{2}\bar{\alpha}_{s}+G_{23}L^{3}\bar{\alpha}_{s}^{2}+G_{34}L^{4}\bar{\alpha}_{s}^{3}+\dots\;\textrm{(LL)\,,}\nonumber\\
g_{2}\left(\alpha_{s}L\right)&=&\,G_{11}L\,\bar{\alpha}_{s}+G_{22}L^{2}\bar{\alpha}_{s}^{2}+G_{33}L^{3}\bar{\alpha}_{s}^{3}+\dots\;\textrm{(NLL)\,.}
\end{eqnarray}
Table~\ref{tab:logs} shows the logarithmic terms present up to the
third order in perturbation theory. At the fixed order level the LL
are term of the form $\alpha_{s}^{n}L^{n+1}$, the NLL those which
goes like $\alpha_{s}^{n}L^{n}$, and so on. Notice that this can be
read off the expansion (\ref{eq:gexpand}) of the exponentiated
resummation functions.

Closed analytic forms for functions $g_1(\alpha_s L)$, $g_2(\alpha_s
L)$ are available for $\tau$ and $\rho$~\cite{resumt}, $B_W$ and
$B_T$~\cite{resumbwbt,resumbwbtrecoil}, $C$~\cite{resumc} and
$Y_3$~\cite{resumy3a}, and are collected in the appendix
of~\cite{ourpaper}. Recently also $g_{3}\left(\alpha_{s}L\right)$ and $g_{4}\left(\alpha_{s}L\right)$
were computed for $\tau$ using effective field theory
methods~\cite{scetthrust}.

\section{MATCHING OF FIXED ORDER AND RESUMMED CALCULATIONS}
To obtain a reliable description of the event shape distributions
over a wide range in $y$, it is mandatory to combine fixed order and
resummed predictions. The two predictions have to be matched in a
way that avoids the double counting of terms present in both. At
NLLA the the expression which has to be matched with fixed NNLO is
given by
\begin{equation}\label{eq:rmatching}
\begin{split}
R\left(y\right)&=\,\left(1+C_{1}\alpha_{s}+C_{2}\alpha_{s}^{2}+C_{3}\alpha_{s}^{3}\right)e^{L\,g_{1}\left(\alpha_{s}L\right)+\,g_{2}\left(\alpha_{s}L\right)+\bar{\alpha}_{s}^{2}G_{21}L+\bar{\alpha}_{s}^{3}G_{32}L^{2}+\bar{\alpha}_{s}^{3}G_{31}L}+\,D\left(y\right)\,,\\
&=\,C\left(\alpha_{s}\right)\Sigma\left(y\right)+\,D\left(y\right)\,,
\end{split}
\end{equation}
where $\Sigma\left(\alpha_{s}\right)$ is the exponentiated part
containing the resummed logarithms, $C\left(\alpha_{s}\right)$ is a
constant and $D\left(\alpha_{s}\right)$ is a remainder functions
which tends to zero as $y\to0$.

A number of different matching procedures have been proposed in the
literature, see for example~\cite{hasko} for a review. In the
so-called $R$-matching scheme, the two expression for
$R\left(y\right)$ are matched. In this case all the coefficients
($C_{2}$, $C_{3}$, $G_{21}$, $G_{32}$ and $G_{31}$) appearing in
(\ref{eq:rmatching}) have to be extracted numerically from the
distributions at fixed order. The increasing number of logarithms present in the fixed order coefficient functions of the NNLO distributions causes large errors on these coefficients.
For this reason we computed the
matching in the so-called $\ln\, R$-matching~\cite{resumall} since
in this particular scheme, all matching coefficients can be
extracted analytically from the resummed calculation. The $\ln\,
R$-matching at NLO is described in detail in~\cite{resumall}. In the
$\ln\, R$-matching scheme, the NLLA+NNLO expression is
\begin{eqnarray}\label{logRmatching}
\ln\left(R\left(y,\alpha_{s}\right)\right)&=&L\,g_{1}\left(\alpha_{s}L\right)\,+\,g_{2}\left(\alpha_{s}L\right)+\,\bar{\alpha}_{S}\left(\mathcal{A}\left(y\right)-G_{11}L-G_{12}L^{2}\right)+{}\nonumber\\
&&+\,\bar{\alpha}_{S}^{2}\left(\mathcal{B}\left(y\right)-\frac{1}{2}\mathcal{A}^{2}\left(y\right)-G_{22}L^{2}-G_{23}L^{3}\right){}\nonumber\\
&&+\,\bar{\alpha}_{S}^{3}\left(\mathcal{C}\left(y\right)-\mathcal{A}\left(y\right)\mathcal{B}\left(y\right)+\frac{1}{3}\mathcal{A}^{3}\left(y\right)-G_{33}L^{3}-G_{34}L^{4}\right)\;.
\end{eqnarray}
The matching coefficients appearing in this expression can be
obtained from (\ref{eq:gexpand}) and are listed in~\cite{ourpaper}.
To ensure the vanishing of the matched expression at the kinematical
boundary $y_{\textrm{\tiny{max}}}$ a further shift of the logarithm
is made~\cite{hasko}.

The renormalisation scale dependence of (\ref{logRmatching}) is
given by making the following replacements:
\begin{eqnarray*}
\alpha_s & \to & \alpha_s(\mu)\;, \\
\mathcal{B}\left(y\right) &\to &
\mathcal{B}\left(y,\mu\right)=2\,\beta_{0}\, \ln x_\mu \,
\mathcal{A}\left(y\right)
+\mathcal{B}\left(y\right)\;,\nonumber \\
\mathcal{C}\left(y\right) & \to &
\mathcal{C}\left(y,\mu\right)=\left(2\,\beta_{0}\, \ln x_\mu
\right)^{2}\mathcal{A}\left(y\right) +2\,\ln x_\mu
\,\left[2\,\beta_{0}
\mathcal{B}\left(y\right)+2\,\beta_{1}\,\mathcal{A}\left(y\right)\right]
+\mathcal{C}\left(y\right)\;,
\label{fixedorderrenscaledependence}\\
g_2\left(\alpha_{s}L\right) &\to &
{g}_{2}\left(\alpha_{s}L,\mu^{2}\right)
=g_{2}\left(\alpha_{s}L\right)+\frac{\beta_{0}}{\pi}
\left(\alpha_{s}L\right)^{2}\, g_{1}'\left(\alpha_{s}L\right)\,\ln
x_\mu \;,
\label{g2mudep} \\
G_{22}&\to & G_{22}\left(\mu\right)=G_{22}\,+\,2\beta_{0}G_{12}\ln
x_\mu
\;,\nonumber\\
G_{33}&\to & {G}_{33}\left(\mu\right)=G_{33}\, +\,4 \beta_{0}
G_{23}\ln x_\mu\,. \label{Gijdeponrenorm}
\end{eqnarray*}
In the above, $g_1'$ denotes the derivative of $g_1$ with respect to
its argument. The LO coefficient ${\cal A}$ and the LL resummation
function $g_1$, as well as the matching coefficients $G_{i\,i+1}$
remain independent on $\mu$.

\section{MATCHED DISTRIBUTIONS AND DISCUSSION}
For the resulting plots of the matched distributions we refer to~\cite{ourpaper}. The most striking observation is that the difference between
NLLA+NNLO and NNLO is largely restricted to the two-jet region,
while NLLA+NLO and NLO differ in normalisation throughout the full
kinematical range. This behavior may serve as a first indication for
the numerical smallness of corrections beyond NNLO in the three-jet
region. In the approach to the two-jet region, the NLLA+NLO and
NLLA+NNLO predictions agree by construction, since the matching
suppresses any fixed order terms. Although not so visible on these plots, the difference between NLLA+NNLO and NLLA+NLO
is only moderate in the three-jet region. The renormalisation scale
uncertainty in the three-jet region is reduced by 20-40\% between
NLLA+NLO and NLLA+NNLO. This effect is due to the smaller renormalization scale dependence of the NNLO contributions. It is also important to observe that the scale dependence remains the same and is larger in the two-jet region, because the resummed calculations at NLLA take into account only the one-loop running of the coupling constant. This has important consequences in the determination of $\alpha_{s}$ and we will comment more on this in the next section.

The description of the hadron-level data improves
between parton-level NLLA+NLO and parton-level NLLA+NNLO, especially
in the three-jet region. The behavior in the two-jet region is
described better by the resummed predictions than by the fixed order
NNLO, although the agreement is far from perfect. This discrepancy
can in part be attributed to missing higher order logarithmic corrections and in part to non-perturbative corrections, which
become large in the approach to the two-jet limit.

\section{CONCLUSIONS AND OUTLOOK}

After the extraction of $\alpha_{s}$ using only the NNLO distributions and the experimental data of ALEPH~\cite{ouralphas}, a new extraction of $\alpha_s$ using the new matched results was performed~\cite{alphasfitmunich} using JADE data. The improvement in the error coming from the inclusion of resummed calculation is not as drammatic as passing from NLO to NLLA+NLO calculations.
As already anticipated, this is due to the fact that the NNLO coefficients compensate the two-loop
renormalization scale variation, whereas the NLLA part only
compensates the one-loop variation.
A further improvement is possible by including the NNLL corrections into the calculations. These corrections are known only for $\tau$, where higher order logarithmic corrections have been computed~\cite{scetthrust} using soft-collinear effective theory (SCET). From these calculations one can extract the functions $g_{3}\left(\alpha_{s}L\right)$ and $g_{4}\left(\alpha_{s}L\right)$. The next step towards the further improvement in the extraction of $\alpha_{s}$ from event-shape distributions could be to compute them for all six observables mentioned here. As shown in~\cite{scetthrust} the subleading logarithmic corrections can also account for
about half of the discrepancy between parton-level theoretical predictions and hadron-level experimental data.

Improvements can also come from non-perturbative corrections. A very recent non-perturbative study for $\tau$ using a low-scale effective coupling~\cite{nonperturbativecorr} shows that
non-perturbative $1/Q$ power corrections cause a shift in the distributions, which can account for an important part of
the difference between parton-level distributions and hadron-level experimental data discussed in the previous section.



\begin{acknowledgments}
We wish to thank the Swiss National Science
Foundation (SNF) which supported this work under contract 200020-117602.
\end{acknowledgments}


\end{document}